\documentclass[epj]{svjour}
\usepackage{graphics}
\usepackage{amsmath}
\newcommand{\bp}{{\bf p}}
\newcommand{\bpp}{{\bf p'}}
\newcommand{\bpz}{{\bf p''}}

\newcommand{\bera}{\langle}
\newcommand{\ket}{\rangle}
\newcommand{\bq}{{\bf q}}

\begin{document}
\title{Calculation of Relativistic Nucleon-Nucleon Potentials in Three-Dimensions}

\author{M. R. Hadizadeh\inst{1,2}\thanks{hadizadm@ohio.edu} \and M. Radin\inst{3}\thanks{radin@kntu.ac.ir}% etc
% \thanks is optional - remove next line if not needed
%\thanks{\emph{Present address:} Insert the address here if needed}%
}                     % Do not remove
%
%\offprints{}          % Insert a name or remove this line
%
\institute{Institute of Nuclear and Particle Physics and Department of Physics and Astronomy, Ohio University, Athens, OH 45701, USA,
\and
College of Science and Engineering, Central State University, Wilberforce, OH 45384, USA,
\and
Department of Physics, K. N. Toosi University of Technology, P.O.Box 16315--1618, Tehran, Iran.}
\date{Received: date / Revised version: date}
% The correct dates will be entered by Springer
%
\abstract{
In this paper, we have applied a three-dimensional approach for calculation of the relativistic nucleon-nucleon potential. 
The quadratic operator relation between the non-relativistic and the relativistic nucleon-nucleon interactions is formulated as a function of relative two-nucleon momentum vectors, which leads to a three-dimensional integral equation.
The integral equation is solved by the iteration method, and the matrix elements of the relativistic potential are calculated from non-relativistic ones. Spin-independent Malfliet-Tjon potential is employed in the numerical calculations, and the numerical tests indicate that the two-nucleon observables calculated by the relativistic potential are preserved with high accuracy. 
\PACS{
      {21.45-v}{Few-body systems}   \and
      {21.45.Bc}{Two-nucleon system} \and
      {24.10.Jv}{Relativistic models}
     } % end of PACS codes
} %end of abstract
\maketitle
\section{Introduction}

The inputs for the relativistic three-body (3B) bound and scattering state calculations \cite{Witala_PRC83}-\cite{Kamada_MPLA24} are the fully off--shell relativistic nucleon--nucleon ($NN$) $t-$matrices, which can be obtained by solving the
relativistic Lippmann--Schwinger (LS) integral equation using relativistic $NN$ interactions.

It is known that there is a nonlinear operator relation between the non-relativistic and the relativistic $NN$ interactions.
So, the first step toward the calculation of relativistic $t-$matrices is the calculation of the relativistic potentials from non-relativistic ones.
To this aim, the matrix elements of the relativistic $NN$ potential in momentum space are traditionally calculated by solving the nonlinear equation using the following different methods.

In the spectral expansion method, the quadratic equation is solved by
inserting a completeness relation of the $NN$ bound and scattering states into the right side of the quadratic equation and by projecting the result into the momentum space \cite{Kamada_PRC66,Gloeckle_PRC33}.
So, by having the non-relativistic potential one can first calculate the $NN$ bound state wave function and scattering half-shell $t-$matrix and used the result to solve the nonlinear equation.

In the iteration method, the nonlinear equation is solved by iteration.
Kamada and Gl\"{o}ckle introduced a powerful numerical technique to calculate the matrix elements of the relativistic $NN$ potential directly from the matrix elements of the non-relativistic $NN$ potential \cite{Kamada_PLB655}. In this method, the nonlinear integral equation is solved using the iteration method to get relativistic and boosted potentials from non-relativistic ones.
It is successfully implemented in the $NN$ problem, but it has not yet been extended to a three-dimensional (3D) approach.

Another method %to calculate the relativistic potential from the non-relativistic one
is to multiply the non-relativistic potential by a function that depends on $NN$ relative momenta, in such a way that both the non-relativistic and the relativistic potentials leads to same phase shifts and observables \cite{Kamada_MPLA80}.
The function is defined in such a way that it changes the non-relativistic kinetic energy to relativistic kinetic energy by rescaling the momentum variables, which leads to the same $2N$ binding energy for both non-relativistic and relativistic potentials.

% 3D
In the past decade a 3D approach based on
momentum vector variables was developed to study the
few--body bound and scattering problems \cite{Hadizadeh_EPJC113_rel}-\cite{Fachruddin_PRC63}.
In the 3D approach one works directly with vector variables which lead to 3D integral equations, whereas the partial wave (PW) representation in the angular momentum basis leads to coupled equations. % on the angular momentum quantum numbers.
In the PW representation, depending on the energy scale of the problem, one must sum PWs, and consequently at higher energies one needs to consider a larger number of PWs, however the 3D approach automatically contains all PWs and the number of equations is energy independent.

We would like to point out that as Polyzou and Elster have shown one can directly calculate the relativistic $t-$matrix from the non-relativistic one, without needing to solve the nonlinear equation. Consequently, one does not need to solve the LS equation for the embedded $NN$ interaction, and one can calculate the fully off-shell relativistic $t-$matrix by following a two-step process. The first step is to obtain the relativistic right--half--shell (RHS) $t-$matrix from the non-relativistic RHS $t-$matrix by an analytical relation proposed by Coester \textit{et al.} \cite{Coester}. The second step is to calculate the fully--off--shell $t-$matrix from the RHS $t-$matrix by solving a first resolvent equation. Keister \textit{et al.} \cite{Keister} proposed the method and it is implemented for the first time in a 3B scattering calculation \cite{Lin_PRC76} in this way.
% 3B relativistic
Using the direct calculation of the relativistic $t-$matrix from the non-relativistic one, recently the relativistic effects were studied in the 3B binding energy using a 3D scheme \cite{Hadizadeh_EPJC113_rel,Hadizadeh_PRC90}. The relativistic 3B wave function was calculated for the first time, and it was shown that the relativistic effects lead to a reduction of about 3\% in the 3B binding energy for two models of a spin-independent Malfliet-Tjon type potential.
Since the 3D approach automatically considers all PWs, if it works for the bound state, it can also be extended to the scattering problem, independent of the range of energy. The next step is to consider the spin and isospin degrees of freedom and work with realistic $NN$ interactions.

In this work, we have applied the iteration method proposed by Kamada and Gl\"ockle to construct the relativistic $NN$ potential from the non-relativistic Malfliet--Tjon potential in a 3D scheme, without using the PW decomposition.

\section{Three-dimensional formulation of the quadratic operator relation between the relativistic and non-relativistic $NN$ potentials}\label{Vr}

According to Bakamjian and Thomas \cite{Bakamjian_PR92} and Fong and Sucher \cite{Fong_JMP5},
the relativistic $NN$ dynamics is specified in terms of the $NN$ mass operator $h$
\begin{eqnarray} \label{eq.h}
\bera \bp| h | \bpp \ket = \omega(\bp) \, \delta(\bp-\bpp) +
V_r(\bp,\bpp),
\end{eqnarray}
where $\omega(\bp)=2E(\bp)=2\sqrt{m^2+\bp^2}$, $m$ is the mass of the nucleons and $\bp$ is the relative momentum of two nucleons.
The connection between the relativistic and non-relativistic $NN$ potentials, \textit{i.e.} $V_r$ and $V_{nr}$, is defined by the quadratic operator equation \cite{Kamada_PLB655}
\begin{eqnarray} \label{eq.V-v}
V_{nr}=\frac{1}{4m} \Big(\omega(\hat{p})V_r+ V_r
\omega(\hat{p})+V_r^2 \Big).
\end{eqnarray}
The matrix elements of the relativistic potential can be obtained from the non-relativistic $NN$ potentials by the projection of Eq. (\ref{eq.V-v}) into the $NN$ basis states $| \bp \ket$
\begin{eqnarray} \label{eq.v-matrix}
&& \bera \bp | V_r | \bpp \ket + \frac{1}{\omega(\bp)+\omega(\bpp)}
\int d\textbf{p}'' \bera \bp |V_r| \bpz \ket \, \bera \bpz |V_r|
\bpp \ket \nonumber \\ [2mm] &&= \frac{4m \bera \bp |V_{nr}| \bpp
\ket}{\omega(\bp)+\omega(\bpp)}.
\end{eqnarray}
In our study we have followed Kamada and Gl\"{o}ckle's strategy \cite{Kamada_PLB655} to obtain the matrix elements of the relativistic $NN$ potential, \textit{i.e.} $\bera \bp | V_r | \bpp \ket$, directly from the non-relativistic one, \textit{i.e.} $\bera \bp |V_{nr}| \bpp \ket$ without using PW decomposition. Here we discuss the numerical solution of Eq. (\ref{eq.v-matrix}) as a function of the magnitude of the momentum vectors and the angle between them. In our calculations we have used the spin independent Malfliet--Tjon (MT) potential, which is a superposition of short--range repulsive and long--range attractive Yukawa interactions \cite{Malfliet}
\begin{eqnarray}
V_{nr}(\bp,\bpp)=\frac{1}{2\pi^2}\left( \frac{V_R}{\bq^2+\mu_R^2} +
\frac{V_A}{\bq^2+\mu_A^2} \right),
\end{eqnarray}
where $\bq=\bpp-\bp$. The parameters of the MT--I potential are given in Table \ref{table:MT--I-parameters}. In order to obtain the matrix
elements of therelativistic potential, we have solved Eq. (\ref{eq.v-matrix}) by the iteration method. A coordinate system is defined by choosing the relative momentum vector $\bp$ parallel to $z-$axis and vector $\bpp$ in the $x-z$ plane, so that Eq. (\ref{eq.v-matrix}) can be written explicitly as
\begin{table}[hbt]
\caption{Parameters of the Malfliet--Tjon I potential.}
\label{table:MT--I-parameters} \centering
\begin{tabular}{cccc}
\hline\noalign{\smallskip}
  $V_A \, \text{(MeV  fm)}$ & $\mu_A \, (\text{fm}^{-1})$ & $V_R \, \text{(MeV  fm)}$ & $\mu_R \, (\text{fm}^{-1})$  \\
\noalign{\smallskip}\hline\noalign{\smallskip}
 -626.8932 &  1.550   &  1438.7228   &  3.11  \\
\noalign{\smallskip}\hline
\end{tabular}
\end{table}
\begin{eqnarray} \label{eq.v-matrix-coordinate}
&& V_r(p,p',x') + \frac{1}{\omega(p)+\omega(p')} \int_{0}^{\infty}
dp'' p''^2 \int_{-1}^{1} dx'' \int_{0}^{2\pi} d\phi'' \cr  && \times
V_r (p,p'',x'') V_r(p'',p',y)
 = \frac{4m
V_{nr}(p,p',x')}{\omega(p)+\omega(p')},
\end{eqnarray}
where
\begin{eqnarray}
y&=&\hat{\textbf{p}}''\cdot\hat{\textbf{p}}'= x'x'' +
\sqrt{1-x'^2}\sqrt{1-x''^2}\cos\phi'',\cr
 x'&=&\hat{\textbf{p}}'\cdot\hat{\textbf{p}},\cr
x''&=&\hat{\textbf{p}}''\cdot\hat{\textbf{p}}.
\end{eqnarray}
We start the iteration with
\begin{eqnarray}
V_r^{(0)}(p,p',x')=\frac{4m V_{nr}(p,p',x')}{\omega(p)+\omega(p')},
\end{eqnarray}
and stop it when the calculated relativistic potential satisfies Eq. (\ref{eq.v-matrix-coordinate}) with a relative error of
$10^{-6}$ at each set point $(p,p',x')$. To speed up the convergence procedure in solving Eq. (\ref{eq.v-matrix-coordinate}) we can
redefine the relativistic potential in each step of the iteration as a linear combination of the calculated relativistic potential in the last two successive iterations as
\begin{eqnarray} \label{eq.convergence-factor}
&& V_r^{(n)}(p,p',x')  \longrightarrow \frac{\alpha
V_r^{(n)}(p,p',x')+ \beta V_r^{(n-1)}(p,p',x')}{\alpha+\beta}; \cr
&&  n=1,2,...
\end{eqnarray}
Kamada and Gl\"{o}ckle have used $\alpha=\beta=1$ in their calculations for the AV18 potential. Our numerical analysis shows that the larger values of $\alpha$ can lead to faster convergence in the solution of Eq. (\ref{eq.v-matrix-coordinate}). 
In Table \ref{table:convergence-alpha-beta} we have shown the number of iterations to reach convergence in Eq. (\ref{eq.v-matrix-coordinate}) for different values of $\alpha$ and $\beta$. 
It indicates that $\alpha=4$ and $\beta=1$ leads to faster convergence for the calculation of the relativistic potential from the MT--I bare potential.
\begin{table}[hbt]
\caption{The number of iterations $N_{iter}$, to reach the convergence in the solution of equation (\ref{eq.v-matrix-coordinate}) for MT--I potential as a function of averaging parameters $\alpha$ and $\beta$.} \label{table:convergence-alpha-beta} 
\centering
\begin{tabular}{ccc}
\hline\noalign{\smallskip}
 $\alpha$ &  $\beta$ & $N_{iter}$
   \\  \noalign{\smallskip}\hline\noalign{\smallskip}
1 &   0  &  17  \\
1  &   1  &  18  \\
2  &   1  &  12  \\
3  &   1  &  10  \\
4  &  1  &  8  \\
5  &  1  &  10  \\
\noalign{\smallskip}\hline
\end{tabular}
\end{table}

For the discretization of the continuous momentum and angle variables we used the Gauss-Legendre quadrature. For the momentum variables a hyperbolic plus linear mapping is used to cover the integration domain $[0, \infty)$ by the subintervals $[0,p_1]\bigcup\,[p_1,p_2]\bigcup\,[p_2, p_{max}]$
\begin{eqnarray}\label{eq16}
p&=&\frac{1+x}{\frac{1}{p_{1}}+(\frac{2}{p_{2}}-\frac{1}{p_{1}})\,x},\\[2mm]
p&=&\frac{p_{max}-p_{2}}{2}\,x+\frac{p_{max}+p_{2}}{2}.
\end{eqnarray}
The typical values for $p_1$, $p_2$ and $p_{max}$ are $4$, $9$ and $60$ fm$^{-1}$ respectively. In our calculations we have used
$100$ mesh points for the momentum variables, $50$ mesh points for the spherical and $10$ mesh points for the azimuthal angle variables.
In each iteration we needed to interpolate on the angle variable $y$ and to avoid extrapolation we have added the extra points $\pm1$ to the angle mesh points $x'$. In order to save run time and memory in solution of equation (\ref{eq.v-matrix-coordinate}) we have used the symmetry property of the
kernel to calculate the integration over azimuthal angle $\phi''$ on the $[0,\pi/2]$ domain
\begin{eqnarray} \label{eq.phi-integration}
\int_0^{2\pi} d\phi'' \, f  (\cos\phi'') = 2\int_0^{\frac{\pi}{2}}
d\phi'' \biggl [ f  (\cos\phi'') +f (-\cos\phi'') \biggr ].\cr
\end{eqnarray}

\begin{figure*}
\resizebox{1.00\textwidth}{!}
{
  \includegraphics{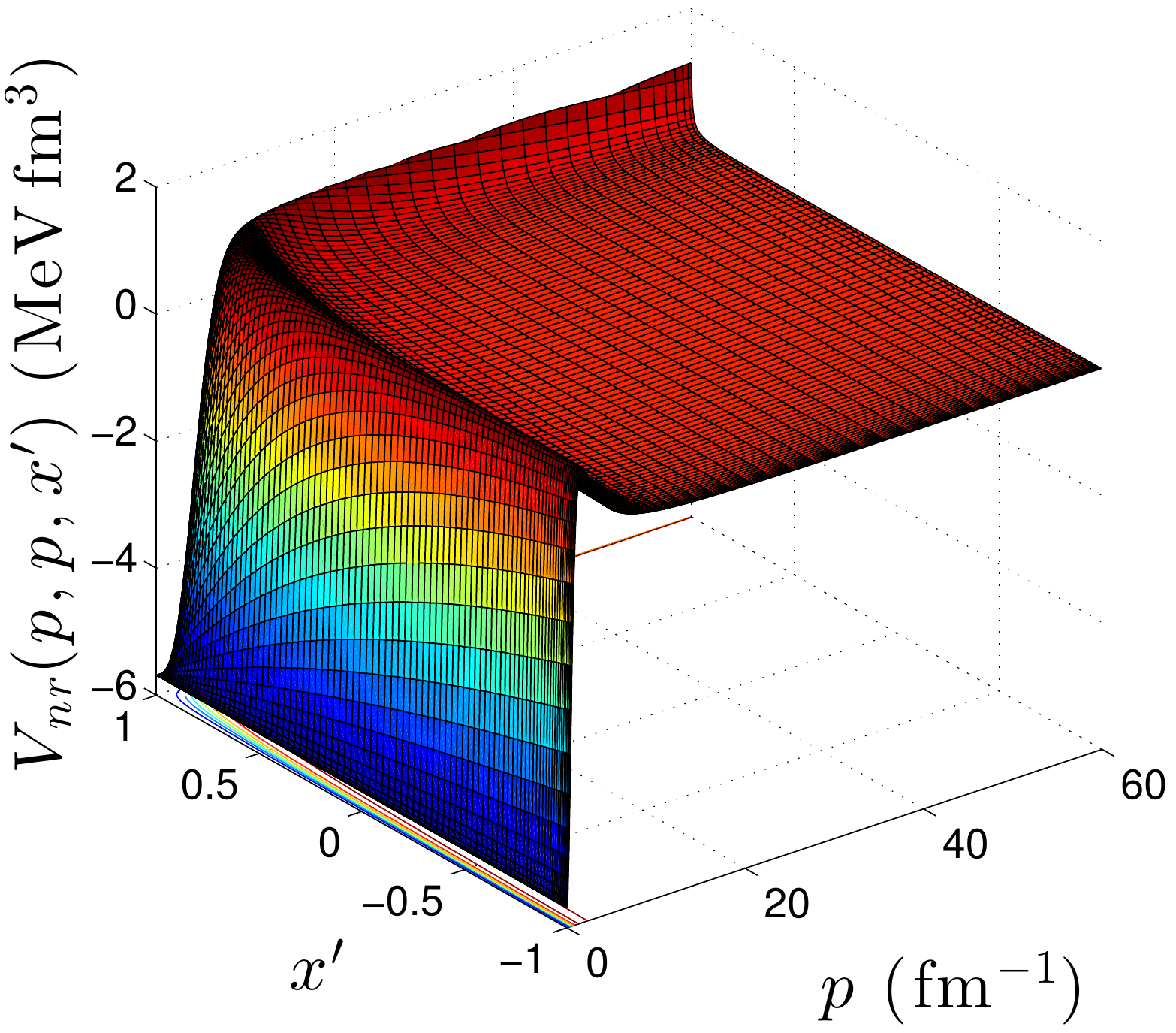}
  \includegraphics{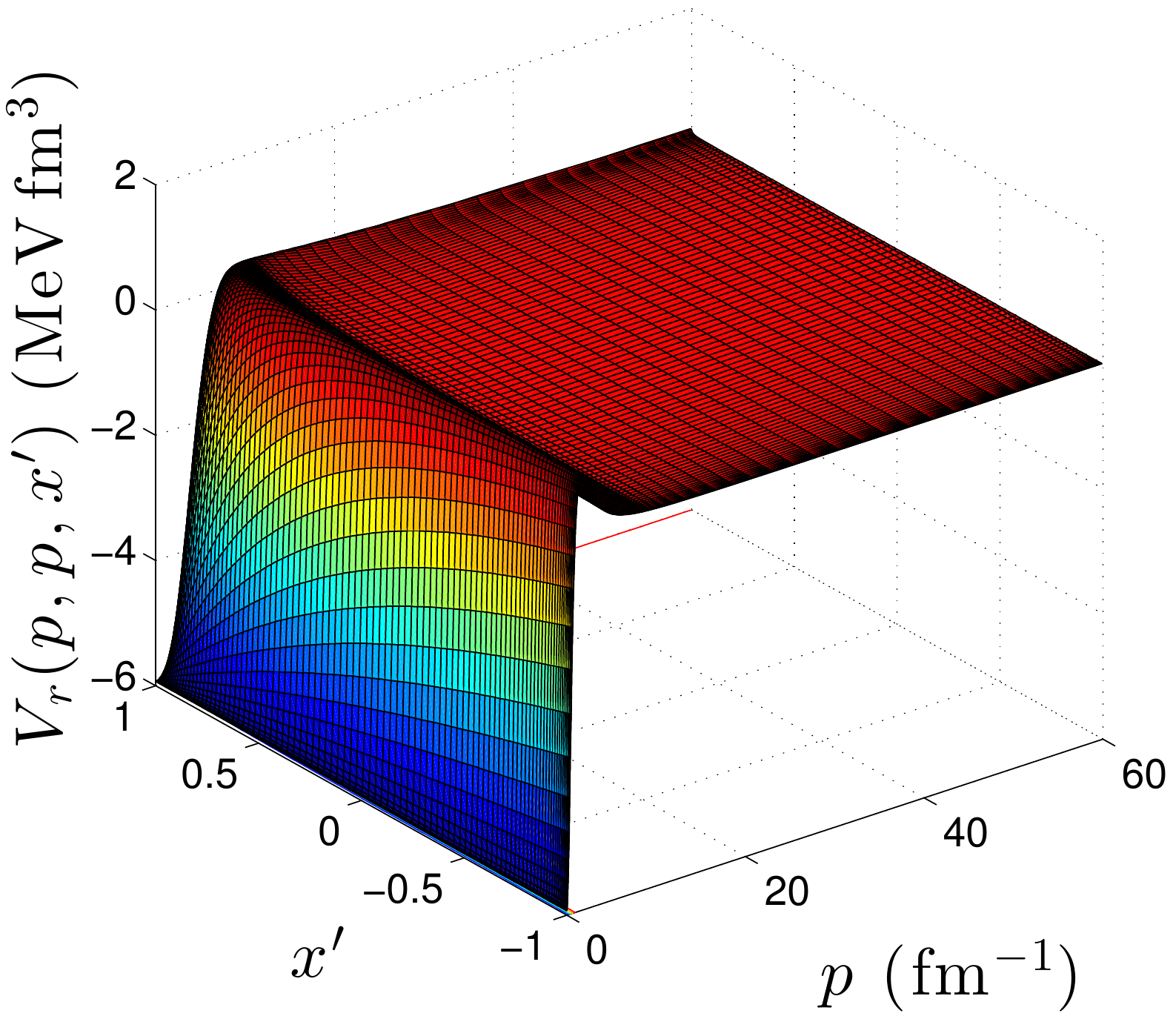}
  \includegraphics{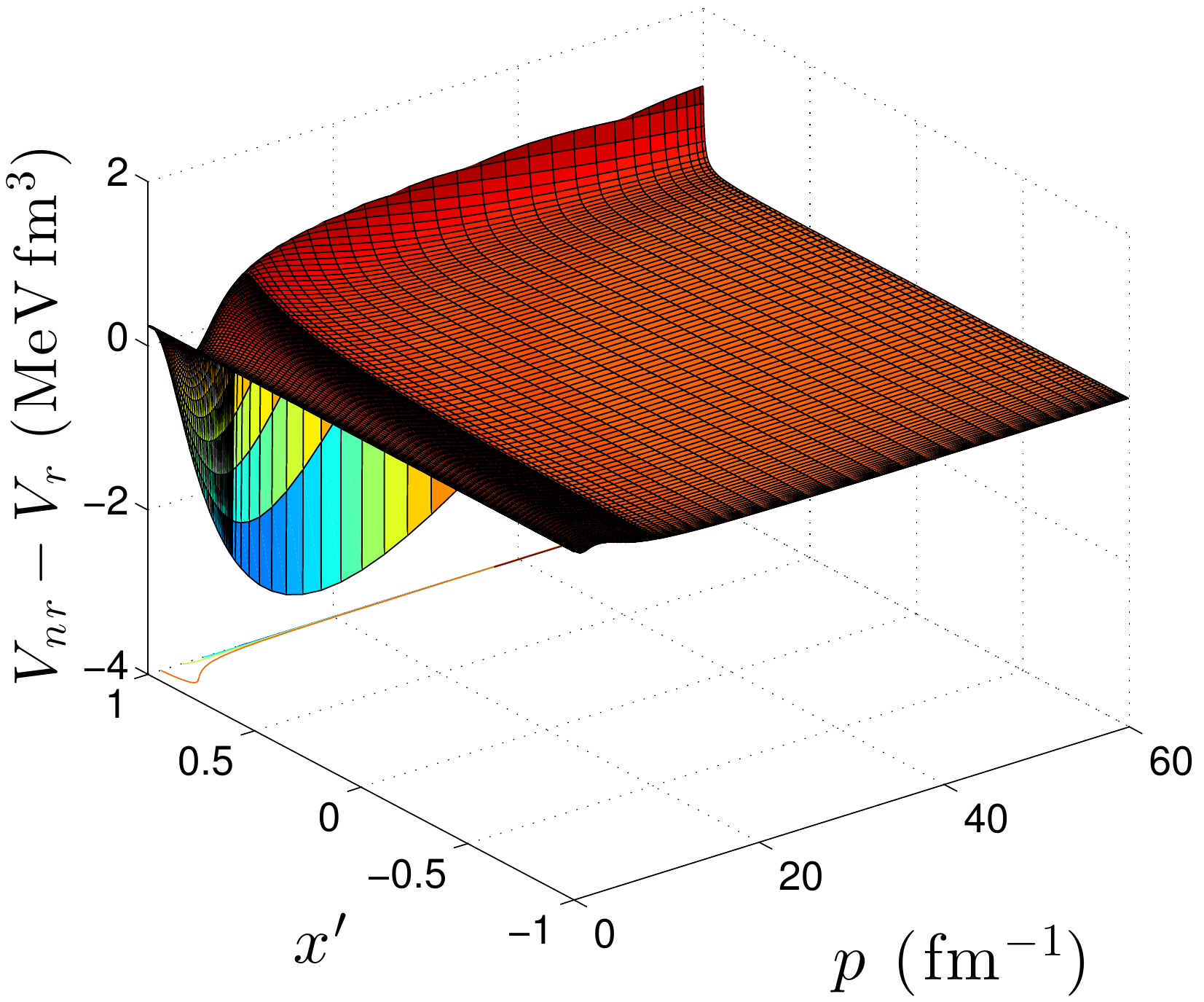}
}
\caption{The matrix elements of the non-relativistic (left panel), the relativistic (middle panel) $NN$ potentials and their differences (right panel) calculated by MT--I potential as a function of 2B relative momenta $p=p'$ and the angle between them $x'$.}
\label{fig:MT--I-px}
\end{figure*}

\begin{figure*}
\resizebox{1.00\textwidth}{!}
{
  \includegraphics{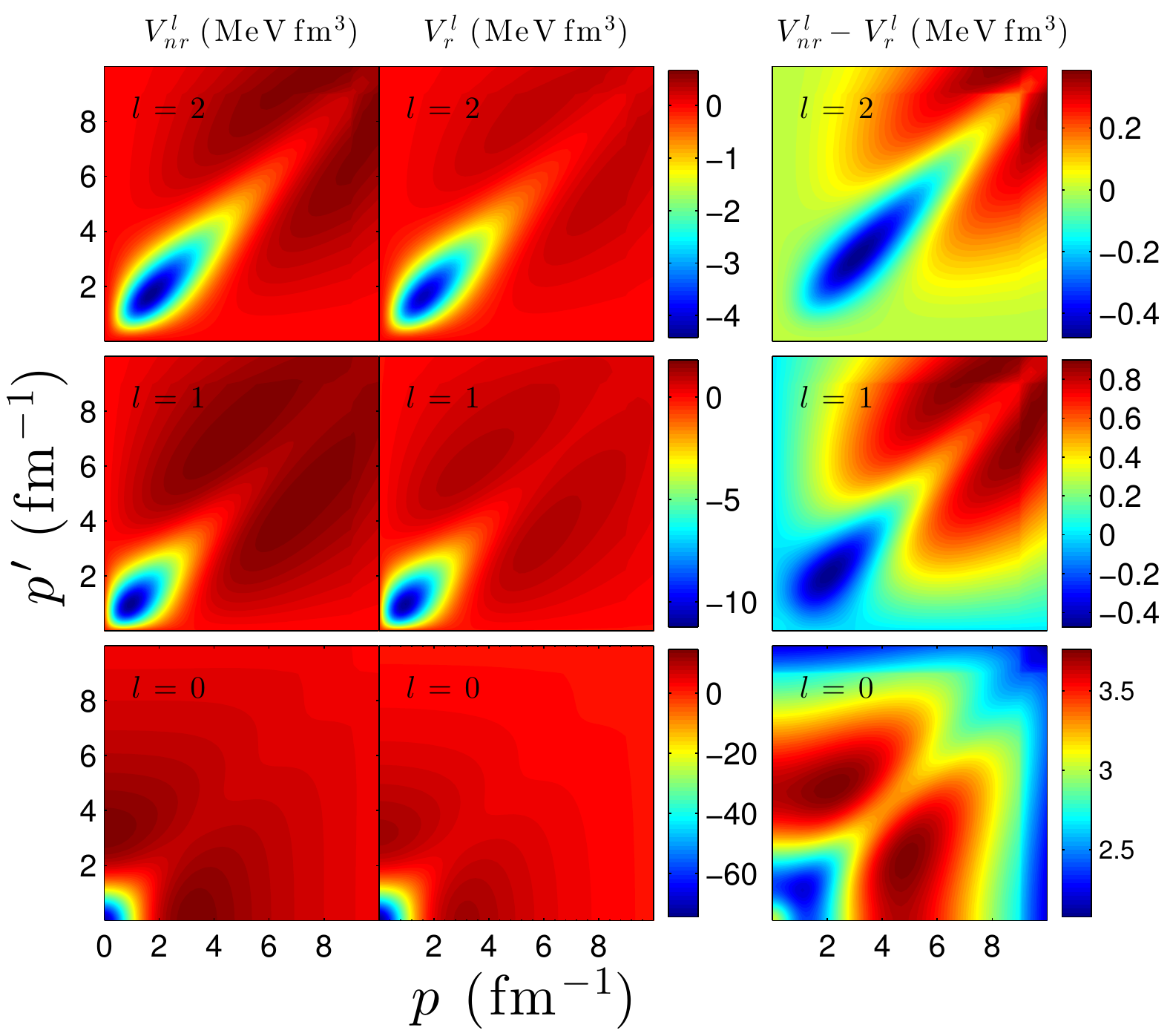}
}
\caption{The matrix elements of the partial wave projection of the non-relativistic (first column), the relativistic (second column) $NN$ potentials and their differences (third column) calculated by MT--I potential as a function of 2B relative momenta $p$ and $p'$. }
\label{fig:MT--I-pp}
\end{figure*}

In Figs. \ref{fig:MT--I-px} and \ref{fig:MT--I-pp} we have shown our numerical results for the relativistic potential calculated from the MT--I potential. The bare MT--I potential as well as the difference between the bare and constructed relativistic potentials is also shown.
The plots of Fig. \ref{fig:MT--I-px} show the non-relativistic and relativistic potentials as well as their difference as a function of the relative momenta $p=p'$ and the angle between them $x'$. It seems the solution of the quadratic equation for the relativistic potential completely changes the structure of the potential at forward angles for diagonal matrix elements $p=p'$, and the relativistic potential is almost smooth in comparison with the non-relativistic potential. 
The corresponding plots in Fig. \ref{fig:MT--I-pp}, show the partial wave projection of the non-relativistic and the relativistic potentials and also their differences, calculated from the 3D representation by $V_l(p,p')=2\pi \int_{-1}^{+1} dx' \, P_l(x') \, V(p,p',x')$, as a function of the relative momenta $p$ and $p'$.
As we can see the matrix elements of the relativistic and non-relativistic potentials are larger for the lower partial waves and consequently their differences become higher.

Table \ref{table:MT--I-convergence1} shows an example of the convergence of the matrix elements of the relativistic potential by iteration number for the fixed points ($p=0.87$ fm$^{-1}$, $p'=2.09$ fm$^{-1}$, $x'=0,\pm1$) for $\alpha=2$ and $\beta=1$.

\begin{table}
\caption{The convergence of the matrix elements of the relativistic potential $V_r(p,p',x')$ (in units of MeV fm$^3$) as a function of iteration number calculated by MT--I bare potential in the fixed points ($p=0.87$ fm$^{-1}$, $p'=2.09$ fm$^{-1}$, $x'=0,\pm1$). The values of the MT--I bare potential $V_{nr}(p,p',x')$ are also given.} \label{table:MT--I-convergence1}
\centering
\begin{tabular}{lccc}
\hline\noalign{\smallskip}
  &  $x'=-1$ & $x'=0$  & $x'=+1$   \\  \noalign{\smallskip}\hline\noalign{\smallskip}
 &  \multicolumn{3}{c}{$V_{nr}(p,p',x')$} \\  \cline{2-4}\noalign{\smallskip}
  & 1.1099096      &  0.7084511     & -1.6327025
   \\\noalign{\smallskip}\hline\noalign{\smallskip}
 Iteration \#    &  \multicolumn{3}{c}{$V_{r}(p,p',x')$} \\ \cline{2-4}
  \noalign{\smallskip} \noalign{\smallskip}
           0 &  1.0525724      &  0.6718530     &  -1.5483583     \\
           1 &  0.7838006      &  0.3920976      & -1.8512322      \\
           2 &  0.8895160      &  0.4979858      & -1.7453002      \\
           3 &  0.8853336      &  0.4938142      & -1.7495059      \\
           4 &  0.8848131      &  0.4932883      & -1.7500527      \\
           5 &  0.8847858      &  0.4932573      & -1.7500929      \\
           6 &  0.8847910      &  0.4932608     &  -1.7500930      \\
           7 &  0.8847932      &  0.4932623      & -1.7500928      \\
           8 &  0.8847938      &  0.4932626     &  -1.7500930      \\
           9 &  0.8847939      &  0.4932626      & -1.7500931      \\
          10 &  0.8847939      & 0.4932626     &  -1.7500932      \\
          11 &  0.8847939      & 0.4932626      & -1.7500933      \\
          12 &  0.8847939      & 0.4932626      & -1.7500933      \\
\noalign{\smallskip}\hline
\end{tabular}
\end{table}

\section{Numerical tests of the relativistic potential}
\subsection{$NN$ bound state }
\begin{figure}
\resizebox{0.48\textwidth}{!}{%
  \includegraphics{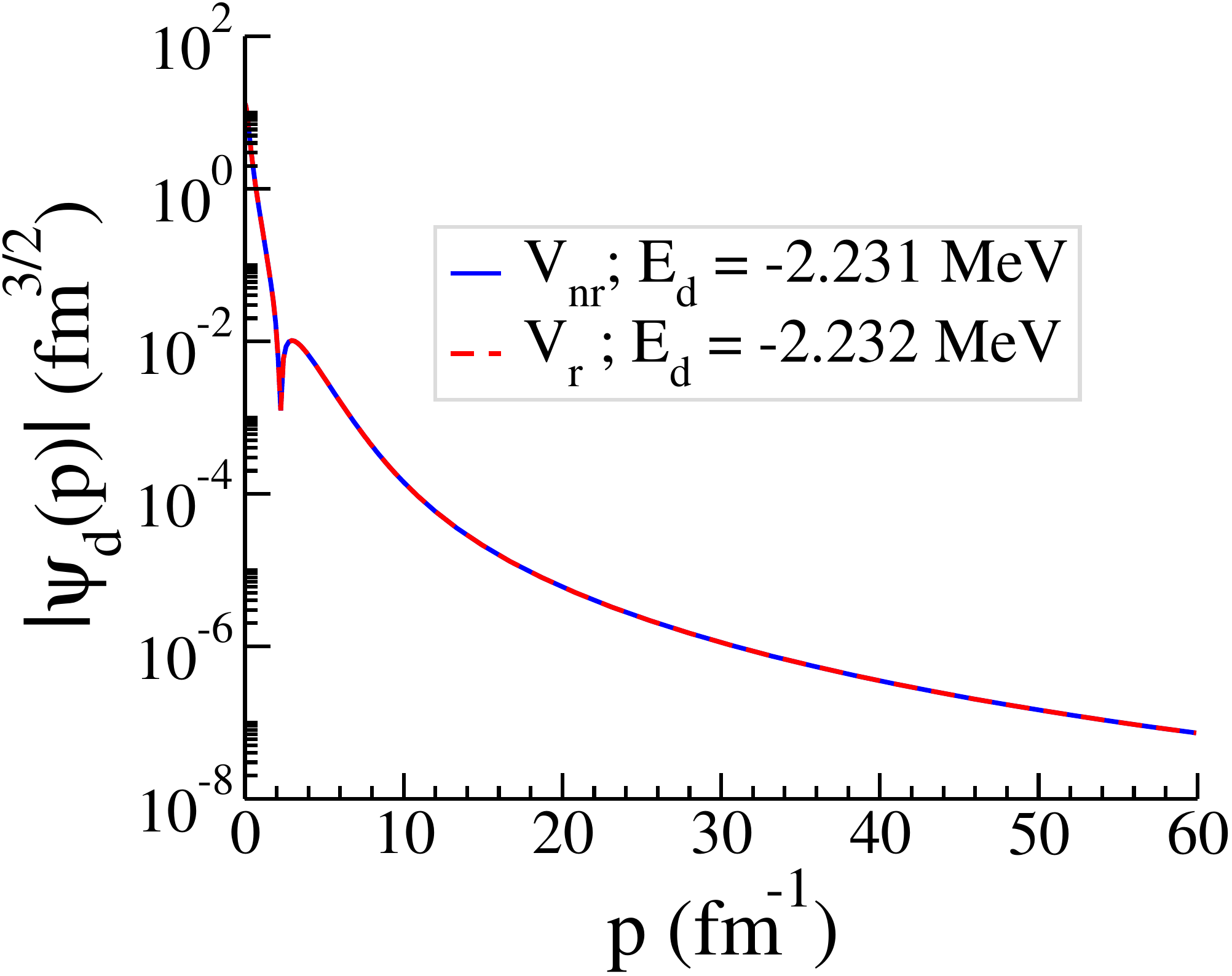}
}
\caption{The deuteron wave function calculated by the MT--I bare and relativistic potentials.}
\label{fig:MT--I_deuteron} 
\end{figure}

The total Hamiltonian of two interacting nucleons in the center of mass system is:
\begin{eqnarray} \label{eq.H}
\bera \bp| H | \bpp \ket =H_0(\bp) \, \delta(\bp-\bpp) +
V_{nr}(\bp,\bpp),
\end{eqnarray}
where $H_0(\bp)=\frac{\bp^2}{m}$ is the free Hamiltonian, $V_{nr}(\bp,\bpp)$
is the non-relativistic $NN$ interaction and $\bp (\bpp)$ is the initial
(final) relative momentum of two nucleons. The Lippmann--Schwinger equation for the two--nucleon bound state is given as
\begin{eqnarray} \label{eq.HLSe}
|\psi_d \ket =\frac{1}{E_d-H_0} V_{nr} |\psi_d \ket,
\end{eqnarray}
which can be represented in momentum space as the following eigenvalue equation
\begin{eqnarray} \label{eq.LSe}
\psi_d (\bp) =\frac{1}{E_d-\frac{\bp^2}{m}} \int d \textbf{p}' \,
V_{nr}(\bp,\bpp) \, \psi_d (\bpp) .
\end{eqnarray}
The relativistic Schr\"{o}dinger equation for the two--nucleon bound state has the form
\begin{eqnarray} \label{eq.Schrodinger-$NN$}
h | \psi_d \ket = M_d | \psi_d \ket,
\end{eqnarray}
where $M_d$ is the deuteron mass. The relativistic deuteron wave function $|\psi_d\ket$ satisfies the eigenvalue equation
\begin{eqnarray} \label{eq.deuteron}
\psi_d(\bp)=\frac{1}{M_d-\omega(\bp)}\int d \textbf{p}' \,
V_r(\bp,\bpp) \, \psi_d(\bpp).
\end{eqnarray}
Our numerical results for the deuteron binding energy and wave function calculated by relativistic and non-relativistic potentials
are given in Table \ref{table:MT--I_deuteron} and Fig. \ref{fig:MT--I_deuteron}.
As we can see the constructed relativistic potential preserves the deuteron binding energy obtained by the bare MT--I potential with high accuracy
and the relative percentage difference of about $0.06$.

\begin{table}
\caption{Deuteron binding energy calculated for MT--I bare and relativistic potentials and their relative difference.}
\label{table:MT--I_deuteron} \centering
\begin{tabular}{ccc}
\hline\noalign{\smallskip}
 $E_d^{nr}$ (MeV) &  $E_d^{r}$ (MeV) &  $ (E_d^{nr}-E_d^{r})/E_d^{nr}\,  \%$
   \\  \noalign{\smallskip}\hline\noalign{\smallskip}
-2.23100    &  -2.23229 & 0.05782 \\
\noalign{\smallskip}\hline
\end{tabular}
\end{table}

\subsection{$NN$ scattering}
The inhomogeneous Lippmann--Schwinger equation which describes two--nucleon scattering can be represented in momentum space as
\begin{eqnarray} \label{eq.T-matrix}
 T_{nr}(\bp,\bpp;E) &=& V_{nr}(\bp,\bpp) \cr &+& \int d \textbf{p}'' \,
\frac{V_{nr}(\bp,\bpz)}{\frac{p_0^2}{m}-\frac{\bpz^2}{m}+i\epsilon}\,
T_{nr}(\bpz,\bpp;E). \quad 
\end{eqnarray}
The differential cross section for elastic $NN$ scattering as a function of incident projectile energy $E_{lab}=2E_{cm}=\frac{2p_0^2}{m}$ is given by
\begin{eqnarray} \label{dsigma}
\frac{d\sigma}{d\Omega} = (2 \pi)^4 \left ( \frac{m}{2} \right )^2 \biggl |
T_{sym}(p_0,p_0,x')  \biggr|^2,
\end{eqnarray}
where
\begin{eqnarray} \label{Ts}
 T_{sym}(p_0,p_0,x') = T_{nr}(p_0,p_0,x') + T_{nr}(p_0,p_0,-x').
\end{eqnarray}
Consequently, the total cross section can be obtained directly from the differential cross section as
\begin{eqnarray} \label{sigma}
\sigma= %\int  d\Omega \, \frac{d\sigma}{d\Omega} =
(2 \pi)^5
\left ( \frac{m}{2} \right )^2  \int _{-1}^{+1} dx' \,  \biggl | T_{sym} \left(p_0,p_0,x' \right)
\biggr|^2.
\end{eqnarray}
The relativistic $NN$ scattering can be described by the relativistic form of the Lippmann--Schwinger equation as
\begin{eqnarray} \label{eq.t-matrix-rel-$NN$}
T_{r}(\bp,\bpp)&=&V_r(\bp,\bpp) \cr &+& \int d
\textbf{p}'' \,
\frac{V_r(\bp,\bpz)}{\omega(p_0)-\omega(p'')+i\epsilon}\,
T_{r}(\bpz,\bpp).
\end{eqnarray}
The relativistic differential and total cross sections can be obtained by Eqs.
(18) and (20) and by replacing $m$ with $\sqrt{m^{2}+p_0^2}$.

In Table \ref{table:sigma}, our numerical results for the total elastic $NN$ scattering cross sections obtained by the constructed relativistic
potential from the MT--I potential are given as a function of the on--shell momentum $p_0$. As we can see the relativistic total cross sections
are in excellent agreement with the corresponding non-relativistic cross sections and have a percentage relative difference of less than $0.007$. $NN$ phase shifts in the PW scheme are calculated by
\begin{eqnarray} \label{phase}
\delta_l(p_0) = \arctan \biggl ( \frac{\texttt{Im }\,
T_l(p_0)}{\texttt{Re} \, T_l(p_0)} \biggr) ,
\end{eqnarray}
where the partial wave $T-$matrix, \textit{i.e.} $T_l(p_0)$, can be obtained from the 3D form of the $T-$matrix, \textit{i.e.} $T(p_0,p_0,x')$, as
 \begin{eqnarray} \label{Tl}
 T_l(p_0) = 2\pi \int_{-1}^{+1} dx' \, P_l(x') \, T(p_0,p_0,x').
\end{eqnarray}
In Table \ref{phase_sp}, we have shown our numerical results for the $s-$ and $p-$wave $NN$ phase shifts as a function of the on--shell momentum $p_0$ calculated from the projection of the 3D form of the non-relativistic and relativistic $T-$matrices by Eq. (\ref{Tl}).
As we can see the relativistic $s-$ and $p-$wave $NN$ phase shifts are in excellent agreement with the corresponding non-relativistic ones and have a relative percentage difference of less than $0.004$ and $0.01$ respectively.

\begin{table}
\caption {The total elastic $NN$ scattering cross section as a function of the on--shell momentum $p_0$ calculated by the MT--I bare and relativistic potentials.}
\label{table:sigma} \centering
\begin{tabular}{lccc}
\hline\noalign{\smallskip}
$p_{0}$ (MeV) &  $\sigma_{nr}$ (mb)  & $\sigma_{r}$ (mb) &  $ |(\sigma_{nr}-\sigma_{r})/\sigma_{nr}  | \%$  \\
\hline\noalign{\smallskip}
 1   &   15274.4     &   15273.3       &  0.00720  \\
  10 &   14526.4      &  14525.5     &  0.00620   \\
   25  & 11485.5      &  11484.9     & 0.00522   \\
   50 &   6367.46      &  6367.31    &   0.00236   \\
   75 &   3432.10      &  3432.08    &  0.00058   \\
   100&   1926.16     &  1926.17     & 0.00052 \\
   200&  297.580   &     297.586     & 0.00202   \\
   300&  129.008  &     129.015     &  0.00543   \\
   400 &  94.6592   &    94.6631    &  0.00412   \\
   500 &   72.6167  &       72.6201     & 0.00468   \\
   600 &    56.9070 &     56.9111    &  0.00720   \\
\noalign{\smallskip}\hline
\end{tabular}
\end{table}

\begin{table}
\caption {The $s-$ and $p-$wave phase shifts, $\delta_0$ and $\delta_1$, calculated by the MT--I bare and relativistic potentials as a function of the on--shell momentum $p_0$.} \label{phase_sp} \centering
\begin{tabular}{lccc}
\hline\noalign{\smallskip}
$p_{0}$ (MeV) &  $\delta_0^{nr}$  & $\delta_0^r$ &  $ |(\delta_0^{nr}-\delta_0^r)/\delta_0^{nr} |\%$  \\
\hline\noalign{\smallskip}
  1 &   178.399202     &    178.399259     & 0.00003   \\
  10 &  164.191097      &   164.191642     & 0.00033   \\
   25  &142.727577      &   142.728682     & 0.00077   \\
   50 & 115.599510      &   115.600883     & 0.00119   \\
   75 &  96.723079      &    96.724553    &  0.00152   \\
   100&  82.604149      &    82.605322     & 0.00142  \\
   200&  46.516325     &     46.517103     & 0.00167   \\
   300&  24.195385     &     24.196576     & 0.00492   \\
   400 &  8.220831       &    8.220032    &  0.00972  \\
   500 &176.141274       &  176.139179     & 0.00119   \\
   600 &166.764703      &   166.759452    & 0.00315
 \\  \hline\noalign{\smallskip}
   $p_{0}$ (MeV)  &  $\delta_1^{nr}$  & $\delta_1^r$ &  $ |(\delta_1^{nr}-\delta_1^r)/\delta_1^{nr} |\%$  \\
\hline\noalign{\smallskip}
  1 &   0.000015228    &    0.000015228    & 0.0017533   \\
  10 &  0.01.520323      &   0.015203706     & 0.0030894   \\
   25  &0.235486179     &   0.235489846               & 0.0015574   \\
   50 & 1.820265835      &   1.820327160               & 0.0033690   \\
   75 &  5.735591957               &    5.735736050      & 0.0025123   \\
   100&  12.07379026       &    12.07410200              & 0.0025819  \\
   200&  35.51818548     &     35.51899747     &           0.0022861   \\
   300&  36.92310114    &     36.92426648                & 0.0031561   \\
   400 &  31.28393516        &    31.28424025    &         0.0009752   \\
   500 &24.51728521       &  24.51716984     &             0.0004706   \\
   600 &18.01488447      &   18.01311044    &              0.0098475   \\
\noalign{\smallskip}\hline
\end{tabular}
\end{table}

\section{Discussion and outlook}
In this paper, we have used a three-dimensional approach to formulating the relativistic nucleon-nucleon potential as a function of the two-body relative momentum vectors.
The quadratic equation which connects the relativistic and non-relativistic nucleon-nucleon interactions is presented in momentum space as a three-dimensional integral equation.
For the first numerical implementation, the integral equation is solved by the spin-independent Malfliet-Tjon potential, and the
matrix elements of the relativistic potential are calculated as a function of the two-body relative momenta and the angle between them. 
Our numerical analysis confirms that the two-body observables calculated from the relativistic potential are preserved. 
The extension of this formalism to realistic nucleon-nucleon interactions with spin degrees of freedom in a momentum-helicity basis state is currently underway.

\section*{Acknowledgements}
We thank Professor Hiroyuki Kamada for helpful discussions and also thank Dr. Jeremy Holtgrave for reading the manuscript in detail and suggesting substantial improvements.
This work is performed under the auspices of the National Science Foundation under Contract No. NSF-HRD-1436702 with Central State University. M. R. H. acknowledges the partial support from the Institute of Nuclear and Particle Physics at Ohio University.

%
% BibTeX users please use
% \bibliographystyle{}
% \bibliography{}
%
% Non-BibTeX users please use

\end{document}